# Synthesis of Core/Shell Ti/TiO$_x$ Photocatalyst via Single-Mode Magnetic Microwave Assisted Direct Oxidation of TiH$_2$

Kunihiko Kato [a], Yunzi Xin [b], Jeongsoo Hong [c], Ken-ichi Katsumata [d], Takashi Shirai [a,b*]

[a] Department of Life Science and Applied Chemistry, Graduate School of Engineering, Nagoya Institute of Technology, Gokiso, Showa-ku, Nagoya, Aichi 466-8555, Japan
[b] Advanced Ceramics Research Center, Nagoya Institute of Technology, Gokiso, Showa-ku, Nagoya, Aichi 466-8555 Japan
[c] Department of Electrical Engineering, Gachon University, 1342 Seongnam-daero, Sujeong-gu, Seongnam-si, Gyeonggi-do, Korea
[d] Photocatalysis International Research Center, Tokyo University of Science, Yamazaki, Noda, Chiba 278-8510 Japan
*Corresponding authors: e-mail: shirai@nitech.ac.jp

## Abstract

Submicron core/shell Ti/TiO$_x$ photocatalyst is successfully synthesized via single-mode magnetic microwave (SMMW) assisted direct oxidation of planetary ball-milled TiH$_2$. The thickness of TiOx shell including highly concentrated defects such as Ti$^{3+}$ and/or oxygen vacancies is controllable in the range from 6 to over 18 nm by varying the treatment time in the SMMW assisted reaction. In addition to its quite narrow optical bandgap (1.34‑2.69 eV) and efficient visible-light absorption capacity, the submicron Ti/TiO$_x$ particle exhibits superior photocatalytic performance towards H$_2$ production from water under both UV and visible-light irradiation to compare with a commercial TiO$_2$ photocatalyst (P-25). Such excellent performance can be achieved by the synergetic effect of enhancement in visible light absorption capacity and photo-excited carrier separation because of the highly concentrated surface defects and the specific Ti/TiO$_x$ core/shell structure, respectively.

## 1. Introduction

Photocatalytic hydrogen (H$_2$) production from water has attracted much interest for several decades [1-3]. Although titanium dioxide (TiO$_2$) is one of the most widely applied photocatalyst, using practically in the limited light region especially UV light due to its wide bandgap (3.0–3.2 eV). Therefore, many researchers have attempted to modify the band structure of TiO$_2$ by introducing donor or acceptor states via doping of metal [4,5] or nonmetal [6-8], which extend optical absorption edge into visible-light region and improve photocatalytic performance. Recently, Ti$^{3+}$ self-doped TiO$_2$ has been paid attention for visible-light photocatalyst since the energy levels induced by Ti$^{3+}$ species and oxygen vacancies can contribute an enhancement of visible-light absorption as well as photocatalytic performance [9-11].

We previously developed a novel one-step synthesize Ti$^{3+}$-oxygen vacancy co-doped

$TiO_2$ from metal Ti particle via single-mode magnetic microwave (SMMW) assisted reactions [12,13]. The synthesized $TiO_2$ exhibits excellent photocatalytic performance under visible-light irradiation despite micrometer-sized photocatalyst particles. However, the particle size of obtained $TiO_2$ strongly depends on the morphology of raw metal particle with particle size in the range of several ten micrometer, which finally cause a limit in a further enhancement for the efficient photocatalytic reaction. Here in present research, we demonstrate submicron highly defective $TiO_2$ via SMMW assisted reaction from planetary ball-milled $TiH_2$ particle as an append system of our previous work. The obtained $TiO_2$ consists of core/shell $Ti/TiO_x$ structure with highly concentrated defects as well as quite narrow optical bandgap of 1.34–2.69 eV. Furthermore, the core/shell $Ti/TiO_x$ particles demonstrates superior photocatalytic performance in $H_2$ production under both UV and visible light irradiation to compare with commercial $TiO_2$. The influence of core/shell $Ti/TiO_x$ structure on photocatalytic activity is also considered in this paper.

## 2. Experimental

### 2.1 Mechanochemical (MC) treatment by planetary ball mill

Titanium hydride ($TiH_2$) powder (3N, powder under 45 μm mesh, Kojundo Chemical Laboratory, Japan) is utilized as raw material. MC treatment is performed by planetary ball mill (Pulverisette-5, Fritsch, Germany) using zirconia ball and pot. MC treated powders were prepared under the following conditions: revolution rate, 300 rpm; ball size, ΦX mm (X = 3, 1 and 0.5); milling time, Y min (Y = 15 or 60) (samples are named as MCX-Y). The particle size distributions are measured by a laser diffraction particle size analyzer (Microtrac MT3200II series, NIKKISO, Japan). Particle morphology is observed by field emission scanning electron microscope (FE-SEM; JSM-7000F, JEOL, Japan). The BET specific surface area (SSA) is analyzed form $N_2$ adsorption / desorption isotherms on high precision gas adsorption measurement instrument (BELSORP-max2, Microtrac BEL, Japan) at 77 K.

**2.2 Synthesis of reduced $TiO_2$ via SMMW assisted reaction**

The core/shell $Ti/TiO_x$ particles are synthesized via the SMMW assisted reaction using 2.45 GHz single-mode MW applicator with $TE_{103}$ mode. The MC treated $TiH_2$ powder of 100 mg is put at maximum magnitude of magnetic MW filed. The samples are heated up at fixed MW output of 100 W under the mixture gas of Ar and $O_2$, whose volume ratio is 97/3 (=Ar/$O_2$) for various reaction times (140, 400 and 840 sec, named as MW-A, MW-B and MW-C, respectively). The crystal structure is characterized by X-ray diffraction (XRD) with Cu-Kα (Ultima IV, Rigaku, Japan). The particle morphology is observed by High Resolution Transmittance Electron Microscope (HR-TEM; JEM-ARM200F, JEOL,

Japan) and crystal phased is confirmed by an equipped Selected Area Electron Diffraction (SAED). For the further information about crystal structure and chemical state in particle surface of obtained samples, Raman spectra are measured by Raman spectrometer (NRS-3100, Jasco, Japan). Surface defects in the initial surface of obtained particles are characterized by X-ray photoelectron spectroscopy with Al Kα X-rays radiation (hv = 1486.6 eV) (XPS: M-prove, SSI, USA). Optical property in the range from UV to visible light is analyzed by UV-Vis spectrophotometer (V-7100, Jasco, Japan).

**2.3 Characterization**

In characterization of photocatalytic activity, the photocatalytic test for $H_2$ production is carried out in a 200 mL home-made quartz glass reactor at ambient temperature and atmospheric pressure. The UV and visible light source are a 200 W Hg-Xe lamp (LA-310UV, HAYASHI, Japan) and Xe lamp (LA-251Xe, HAYASHI, Japan) at 1 mW cm$^{-2}$ of light power density, respectively. 25 mg of the sample is added into the 100 ml of mixture solution of distilled water and methanol as sacrificial agent, whose volume ratio is 9 to 1. 1 mL of gas is taken out from the reactor by syringe constant time after the irradiation and analyzed by using a gas chromatograph (GC-2014, Shimadzu) equipped with a thermal conductivity detector (TCD). Photoluminescence (PL) spectra are measured by spectrofluorometer (FP-8500, Jasco, Japan) at an excitation wavelength of 330 nm and

405 nm.

## 3. Results and discussion

### 3.1 Preparation of submicron TiH$_2$ particle by planetary ball mill

The particle size distributions of raw and MC treated TiH$_2$ powder are shown in Figure 1. The average particle size (d$_{50}$) of raw, MC3-15, MC1-15, MC0.5-15 and MC0.5-60 are 25.49, 0.61, 0.87, 0.53, and 0.39 μm, respectively. In comparison of the ball media size, the narrowest distribution is exhibited in MC0.5-15 treated with the smallest ball media (Φ 0.5 mm). In addition, the distribution becomes narrower and d$_{50}$ decreases with increasing treatment time (MC0.5-60). It is expected that applied mechanical energy towards TiH$_2$ particles per ball media increase with larger ball media size [14], resulting in strong adhesion among the generated fine particles at the same time of grinding. Thus, MC3-15 demonstrates the clear bimodal distribution because of both effects of grinding and adhering. In fact, the formation of coarse particles is confirmed from FE-SEM images as shown in Figure 2. Furthermore, MC3-15 exhibits relatively higher SSA despite the presence of the large sized particles as shown in Figure 3, it implies the formation of agglomerates among fine particles as mentioned above. In this time, MC0.5-60 is selected as the target material for the synthesis of TiO$_2$ via the SMMW assisted reaction, since showing the highest SSA among the planetary ball milled TiH$_2$ powders.

**3.2 Characterization of highly defective TiO$_2$ synthesized via SMMW assisted reaction**

Figure 4a shows XRD patterns of the samples synthesized in the various reaction time during SMMW assisted reaction. The obtained materials compose of two kind of crystal phase, which are assigned to TiO$_2$ rutile phase and Ti. It is expected that Ti phase is formed by the dehydrogenation of TiH$_2$ firstly, the TiO$_2$ rutile crystal is grown with increase of reaction time, accompanying with the disappearance of Ti phase. Moreover, the long-range disordered structure in TiO$_2$ rutile crystal is observed more significantly in the sample through short time synthesis, especially MW-A (Figure 4b). Additionally, HR-TEM analysis reveals the specific core/shell structure of the obtained particle as shown in Figure 5. In MW-A which is the short-time synthesis, the core/shell structure with dark/light contrast is clearly observed in Figure 5a-b. The lattice with dark contrast corresponds to Ti (101) since d-spacing is 0.22 nm. Figure 5c shows HR-TEM image of the particle surface of MW-B. A strong black contrast is observed in the interface of the part of core (area A) and shell (area B). Such black contrast might be contributed to defects induced by strain of lattice mismatch between core and shell parts. The d-spacing of 0.27 nm in core and 0.22 nm in shell are confirmed, which are corresponded to $TiO_2$ rutile (101) and Ti (101) planes, respectively. In addition, the selected area electron diffraction (SAED) patterns in core and shell displayed in Figure 5d and 5e have a good agreement with diffraction pattern of rutile $TiO_2$ and metal Ti, respectively. The shell thickness of $TiO_x$ can be identified as 10-15 nm in sample MW-B. On the other hands, MW-C has a larger shell thickness than MW-B, as demonstrated in Figure 5f. The thickness of $TiO_x$ can be estimated as over 18 nm. Additionally, the lattice contrast is observed more clearly and black contrast attributed by $Ti^{3+}$ defects exist in the particle surface. Therefore, for obtained particles compose of Ti core-$TiO_x$ shell structure as the illustration shown in Figure 5g, the shell-thickness of $TiO_x$ is controllable in the range from 6 to over 18 nm by varying the treatment time in the SMMW assisted reaction.

For the further investigation of the chemical bonding in the obtained particle surface, Raman spectroscopic measurement is conducted as shown in Figure 6a. As a result, we can find obvious peaks which are attributed to Raman active modes of rutile crystal [15]. Furthermore, the peak of $E_g$ mode at about 440 cm$^{-1}$ shifts towards lower wavenumber (red-shift) in the synthesis with short reaction time. It is known that $E_g$ mode of rutile is sensitive against oxidation state of $TiO_2$, a significant red–shift can be detected due to defects such as oxygen vacancies in $TiO_2$ [16,17]. In this time, the chemical composition of the prepared $TiO_2$ is estimated by the approximation curve deduced from Raman shift vs chemical composition (O/Ti) diagram in the reported paper [17]. As shown in Figure 6b, the estimated O/Ti increase linearly in the range from 1.85 to 1.94 with increase of the reaction time. Furthermore, the presence of highly concentrated defects in the initial surface of the obtained particles is confirmed by XPS analysis. In the XPS spectra of $Ti_{2p}$ orbital (Figure 7a), the peaks can be decomposed into three components at 458.8, 457.1 and 455.4 eV which are assigned with $Ti^{4+}$, $Ti^{3+}$ and $Ti^{2+}$, respectively [18,19]. Figure 7b summarizes the relation between percentage of Ti valence species ($Ti^{4+}$, $Ti^{3+}$ and $Ti^{2+}$) and the treatment time in SMMW assisted reaction. The concentration of $Ti^{3+}$ and $Ti^{2+}$ significantly decreases with the increase of reaction time. Additionally, a notable difference is observed also in the shape of XPS $O_{1s}$ spectra shown in Figure 7c. The peaks at 530.1, 531.7, and 533.2 eV are assigned with $O^{2-}$ (Ti-O) and $OH^-$ and chemisorbed water molecules, respectively [20,21]. It is known that oxygen vacancies induce dissociation of $H_2O$, then taking place chemisorption as OH group on defective $TiO_2$ surface [22.23]. Thus, these results indicate an evident fact that highly concentrated defects ($Ti^{3+}$ and oxygen vacancies) exist on surface of the obtained particles. Regards to the optical property, the synthesized core/shell $Ti/TiO_x$ particles exhibits the superior visible-light absorption capacity with grey and black coloration as shown in Figure 8a.

This improvement of light absorption is commonly seen in deficient $TiO_2$ [24-26]. Additionally, deficient $TiO_2$ generally shows the localized impurity levels induced by $Ti^{3+}$ or oxygen vacancy at 0.75–1.18 eV below conduction band (CB) minimum [27,28]. The electrons in these levels can transfer to CB by a thermal or photoexcitation process to form the unoccupied states [29]. This process corresponds to the high light absorption in over 600 nm as shown in the synthesized $TiO_2$ with highly concentrated defects. Furthermore, the obtained core/shell $Ti/TiO_x$ exhibit quite narrower optical bandgap to compare with a commercial rutile $TiO_2$ ($\approx$ 3.0 eV). The optical bandgap of MW-A, MW-B and MW-C deduced by Tauc plot (see Figure 8b) are 1.34, 2.34 and 2.69 eV, respectively. The photocatalytic performance of the core/shell $Ti/TiO_x$ particle is characterized by $H_2$ production under UV and visible light irradiation as shown in Figure 9a and 9b, respectively. The obtained $Ti/TiO_x$ showed the appreciable photocatalytic performance under both UV and visible light to compare with highly efficient $TiO_2$ photocatalyst (P-25). Figure 8c shows photocatalytic performance per unit surface which is normalized by the SSA value of the obtained samples. The photocatalytic activity under visible light irradiation increase with increase of the reaction time, not depending on increase of the defect concentration. The obtained core/shell $Ti/TiO_x$ particles demonstrate the photocatalytic activity per unit surface area over ten time higher than P-25 under both UV and visible light irradiation.

### 3.3 Influence of core/shell-Ti/TiO$_x$ structure on photocatalytic performance

Herein, we suggest that the superior photocatalytic performance of the prepared TiO$_2$ can be attributed to synergetic effect of enhancement in visible light absorption capacity and photo-excited carrier separation achieved by the disordered structure with highly concentrated surface defects and the specific Ti core/TiO$_x$ shell structure, respectively as shown in Figure 10. In the interface of metal/n-type semiconductor core/shell structure such as Ti/TiO$_x$, an upward band bending occurs to reach Fermi level ($E_F$) equilibrium between Ti and TiO$_2$ due to the difference in their work function, would result in a formation of a Schottky barrier in the interface [30-33]. The photo-excited electrons in Ti will inject to TiO$_2$ side efficiently, while the photo-excited hole from TiO$_2$ could transfer immediately to Ti side. Meanwhile, the highly concentrated defects (Ti$^{3+}$ species and/or oxygen vacancies) tend to work as electron trap centers, which lead to restrain the recombination between electrons and holes, and increase the lifetime of excited electrons as well [34-36]. Furthermore, the defects can also improve the electric conductivity [37-39] and promote the electron transfer. For further investigation of the photocatalytic mechanism, an additional characterization of excitation-wavelength dependent photoluminescence of the obtained particles is conducted as shown in Figure 11. Under UV light excitation, MW-A shows superior photo-excited carrier separation, whereas the PL intensity significantly increase by visible light excitation (= 405 nm) in the wide range from 450 to 550 nm, where are equivalent to energy gap from 2.25 to 2.76 eV. In this case, it is considered that a localized state 0.75 eV bellow CB minimum in the bandgap of $TiO_2$ induced by oxygen vacancies would involve the photo-excited carrier recombination. Regards to XPS analysis results, MW-A shows highly concentrated oxygen vacancies as displayed in XPS $O_{1s}$ spectra. Although effects of oxygen vacancies for photocatalytic activity is still a matter of controversy [40], excessive amount of oxygen vacancies would serve as recombination centers, affecting a negative impact for photocatalytic performance under visible light irradiation in our case. Moreover, the long-range disordered structure in $TiO_2$ rutile crystal is observed more significantly in the XRD pattern of MW-A, which can also cause a deterioration of photocatalytic activity [41]. On the other hands, UV light excited carrier separation would be strongly affected by the shell thickness of $TiO_x$. The dependence of $TiO_x$ shell-thickness on PL intensity can be seen according to the above described results. As a conclusion, the specific Ti core-$TiO_x$ shell structure would dominantly contribute the efficient photo-excited carrier separation regardless the chemical structure of $TiO_x$ in the case of UV light excitation.

## 4. Conclusions

In conclusion, we have successfully synthesized the submicron core/shell Ti/TiO$_x$ particles via the SMMW assisted direct oxidation of planetary ball milled TiH$_2$. The thickness of TiO$_x$ shell is controllable in the range from 6 to over 18 nm by varying the treatment time in the SMMW assisted reaction. The synthesized Ti/TiO$_x$ particles exhibit the excellent visible light absorption as well as quite narrower optical bandgap of 1.34–2.69 eV, whose surface chemical composition can be controlled in the range from 1.85 to 1.94 according to the estimation from Raman shift. The photocatalytic performance is characterized by H$_2$ production from water, the synthesized Ti/TiO$_x$ particles perform the superior activity to compare with commercial TiO$_2$ (P-25) under both UV and visible light irradiation. Furthermore, the interface of Ti core / TiO$_x$ shell structure contribute to enhance an efficiency of spatial charge separation between photo-excited electrons and holes by a Schottky barrier, results in the further improvement of the photocatalytic performance. The SMMW synthesis method demonstrated in this study provides feasible synthesis routine for functional metal oxide materials with specific chemical and physical properties.

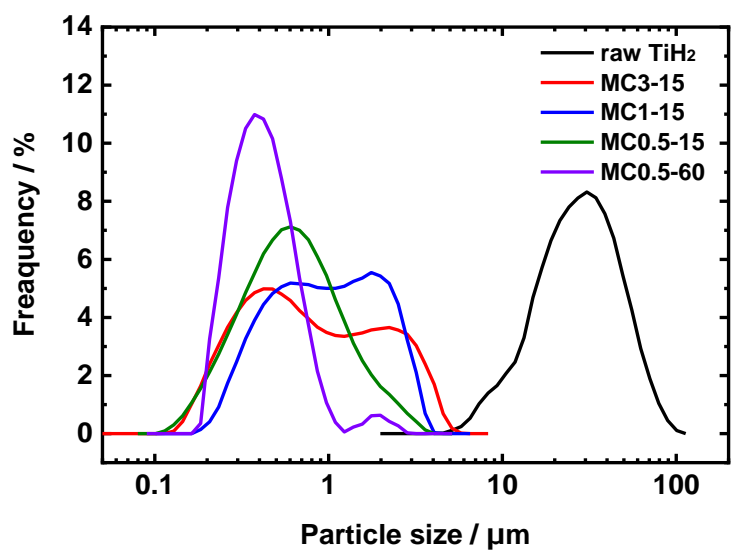

Figure 1. Particle distribution of raw and planetary ball milled $TiH_2$.

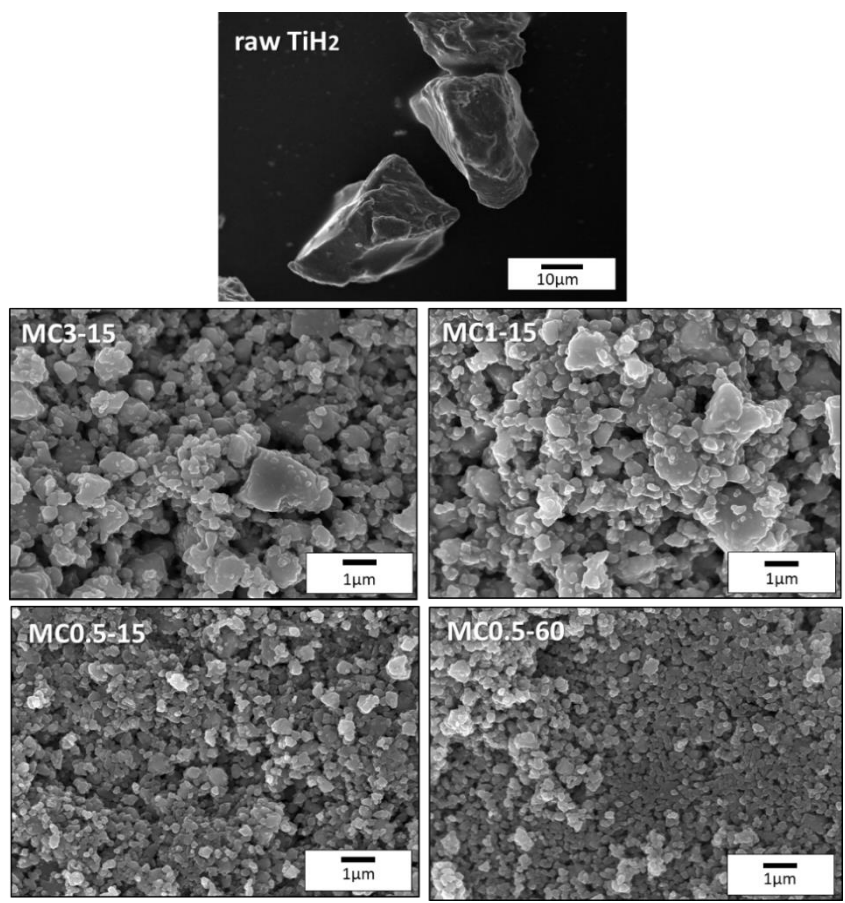

Figure 2. Particle morphology by FE-SEM for raw and planetary ball milled $TiH_2$.

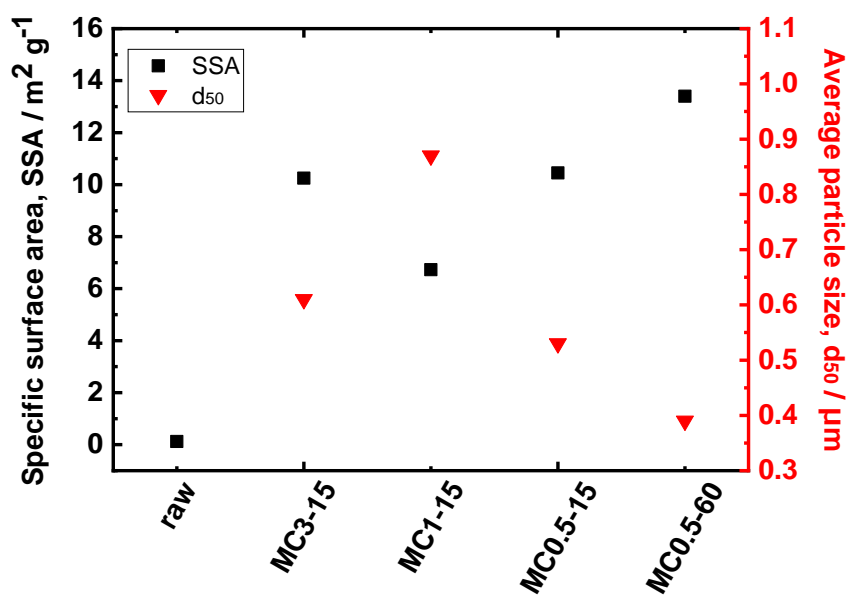

Figure 3. Correlation of BET specific surface area and average particle size deduced from particle distribution.

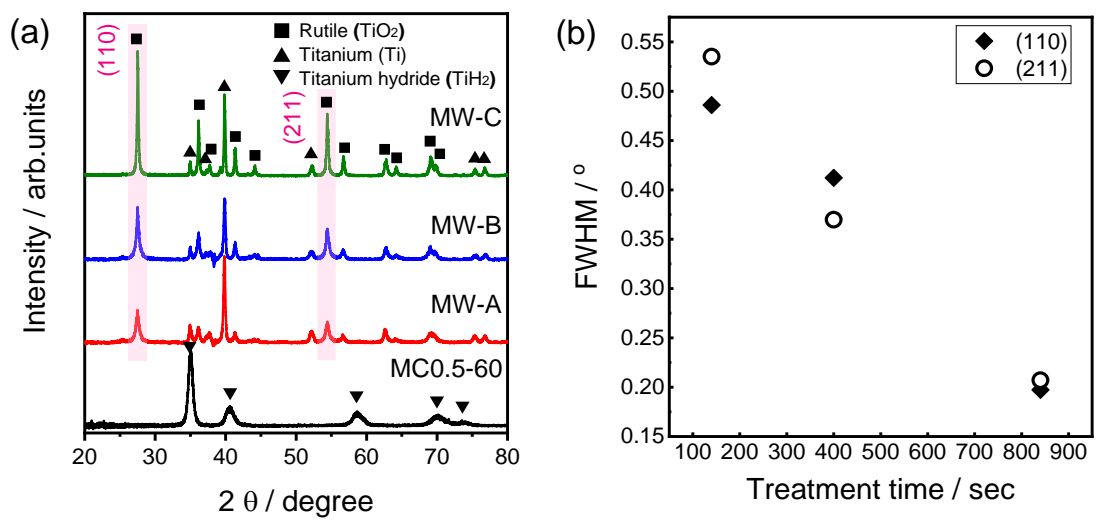

Figure 4. Crystal structure of the planetary ball milled TiH$_2$ and the obtained samples through SMMW assisted reaction: (a) XRD patterns; (b) Full width at half maximum (FWHM) of peaks attributed to TiO$_2$ rutile (110) and (211).

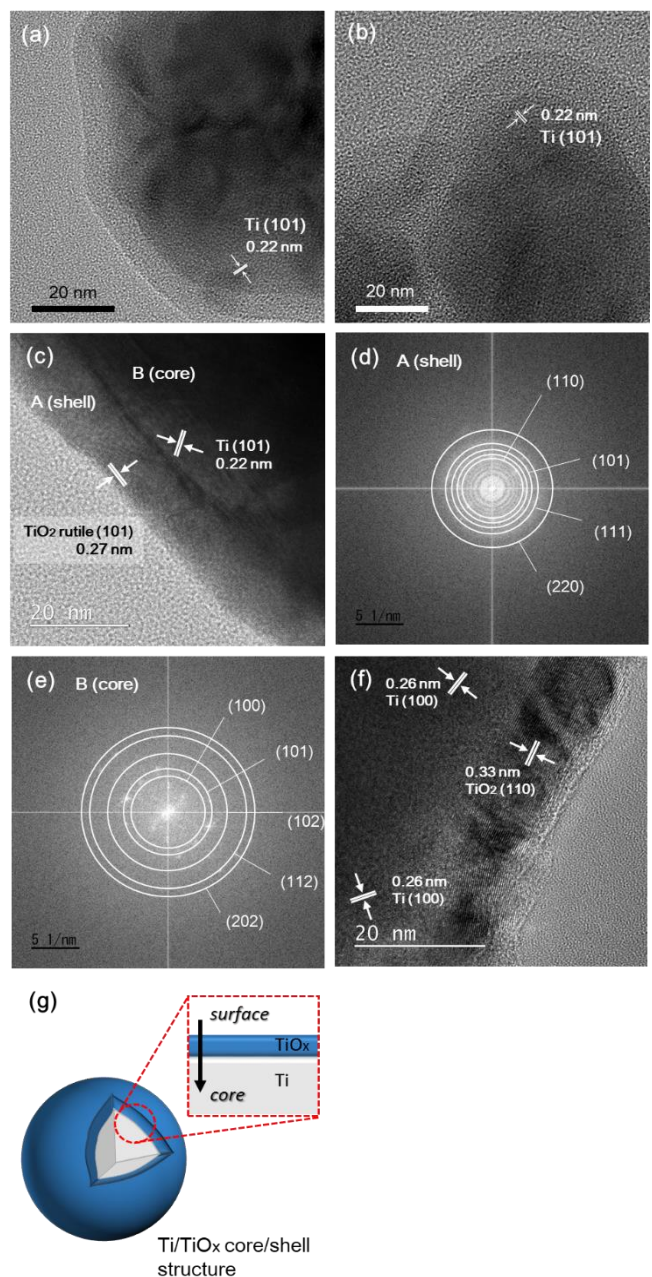

Figure 5. TEM analysis of the obtained TiO$_x$; (a-b) HR-TEM of core/shell structure (MW-A), (c) HR-TEM image of the obtained particle surface (MW-B), (d) SAED pattern in the part of particle shell, (e) SAED pattern in the part of particle core, (f) HR-TEM image of the obtained particle surface (MW-C) (g) Scheme of the obtained particle with Ti/TiO$_x$ core/shell structure via SMMW assisted reaction.

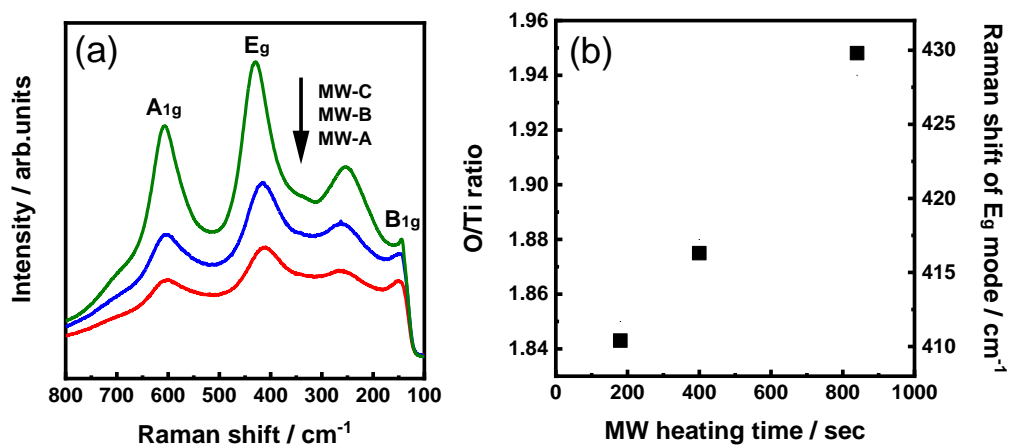

Figure 6. (a) Raman spectra of the synthesized $TiO_2$, (b) Correlation of MW heating time with O/Ti ratio of the synthesized $TiO_2$ calculated by Raman shift of $E_g$ mode.

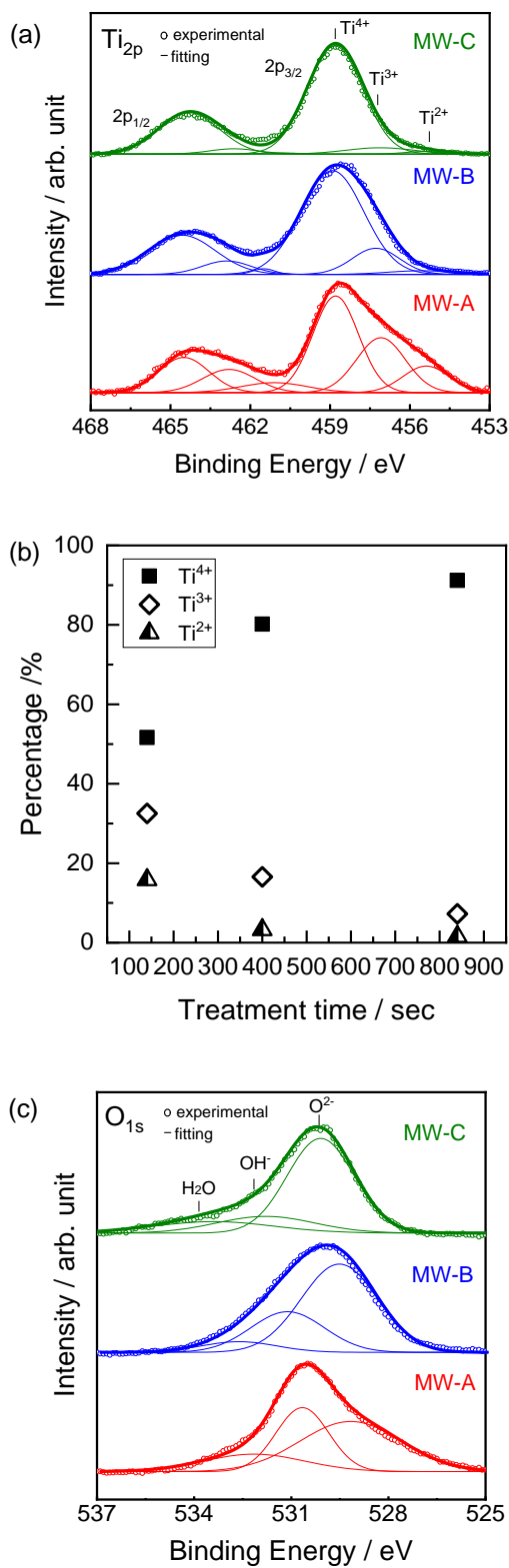

Figure 7. Surface chemical state of the obtained particles: (a) XPS $Ti_{2p}$ spectra, (b) Percentage of Ti valence species ($Ti^{4+}$, $Ti^{3+}$ and $Ti^{2+}$), (c) XPS $O_{1s}$ spectra.

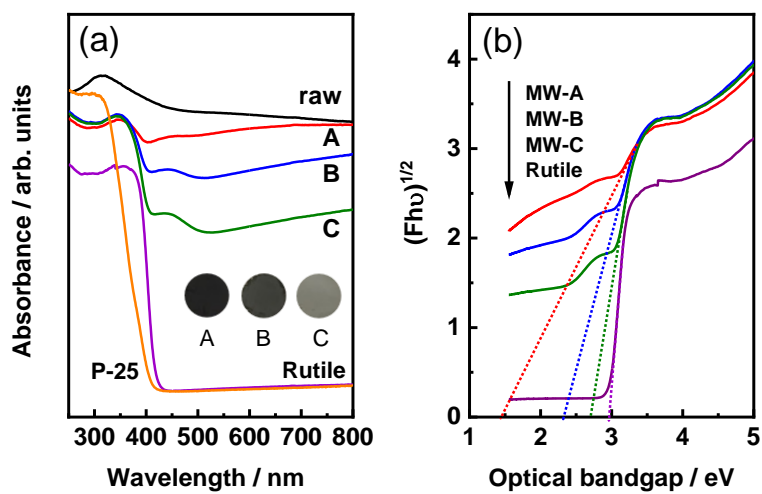

Figure 8. Characterization of optical property of the synthesized submicron TiO$_2$ in UV–vis; (a) Absorption spectra. Inset pictures display apparent color of the obtained particle. (b) Tauc plot.

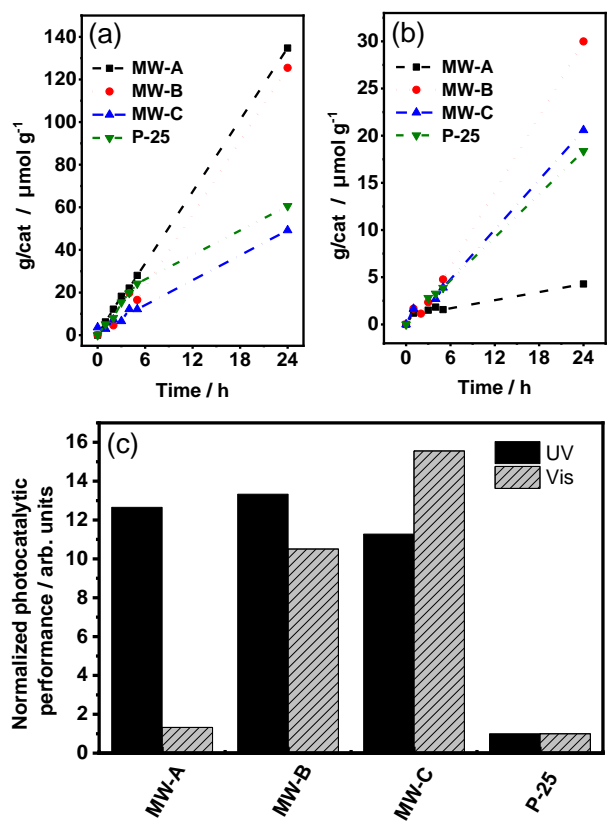

Figure 9. Photocatalytic activity characterized by H$_2$ production under irradiation of (a) UV and (b) visible light. (c) Photocatalytic performance normalized by SSA.

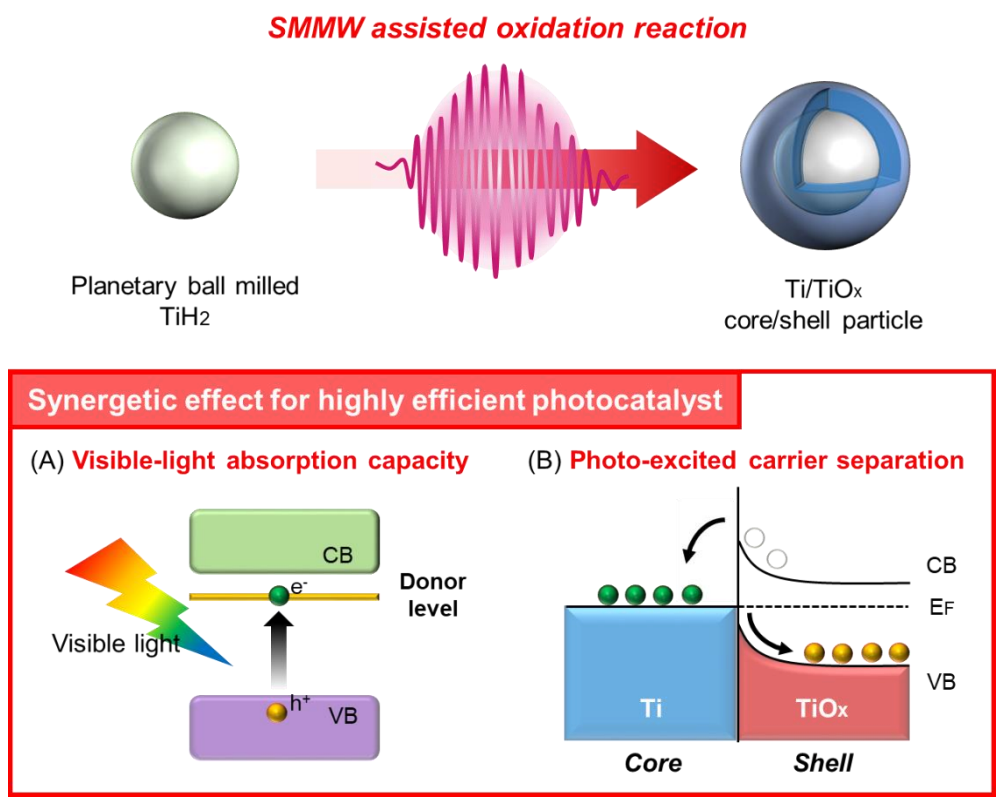

Figure 10. Schematic illustration of highly efficient photocatalyst via SMMW assisted oxidation reaction.

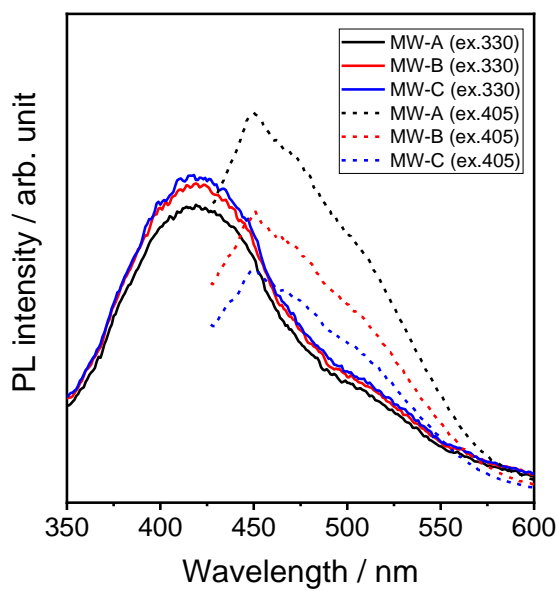

Figure 11. Excitation-wavelength dependent photoluminescence. The solid and dash line represent UV light (330 nm) and visible light (405 nm) excitation, respectively.